# Improving living biomass C-stock loss estimates by combining optical satellite, airborne laser scanning, and NFI data

Johannes Breidenbach *, Norwegian Institute of Bioeconomy Research (NIBIO), National Forest Inventory, Ås, Norway

Janis Ivanovs, Latvian State Forest Research Institute (Silava), Forest operations and energy, Salaspils, Latvia

Annika Kangas, Natural Resources Institute Finland (Luke), Bioeconomy and environment, Joensuu, Finland

Thomas Nord-Larsen, Department of Geosciences and Natural Resource Management, University of Copenhagen, 1958 Frederiksberg C, Denmark

Mats Nilsson, Swedish University of Agricultural Sciences (SLU), Department of Forest Resource Management, Umeå, Sweden

Rasmus Astrup, Norwegian Institute of Bioeconomy Research (NIBIO), Division of Forest and Forest Resources, Ås, Norway

* Contact: [first_name].[last_name][at]nibo.no

## 1 Abstract

Policy measures and management decisions aiming at enhancing the role of forests in mitigating climate-change require reliable estimates of C-stock dynamics in greenhouse gas inventories (GHGIs). The aim of this study was to assemble design-based estimators to provide estimates relevant for GHGIs using national forest inventory (NFI) data. We improve basic expansion (BE) estimators of living-biomass C-stock loss using field-data only, by leveraging with remotely-sensed auxiliary data in model-assisted (MA) estimators. Our case studies from Norway, Sweden, Denmark, and Latvia covered an area of >70 Mha. Landsat-based Forest Cover Loss (FCL) and one-time wall-to-wall airborne laser scanning (ALS) data served as auxiliary data. ALS provided information on the C-stock before a potential disturbance indicated by FCL. The use of FCL in MA estimators resulted in considerable efficiency gains, which in most cases were further increased by using ALS in addition. A doubling of efficiency was possible for national estimates and even larger efficiencies were observed at the sub-national level. Average annual estimates were considerably more precise than pooled estimates using NFI data from all years at once. The combination of remotely-sensed and NFI field data yields reliable estimators which is not necessarily the case when using remotely-sensed data without reference observations.

# 2  Introduction

The struggle to achieve a "well-below" 2° Celsius mean temperature increase by 2100 as concerted in the Paris agreement (UN 2015), requires policy measures to fulfill emission reduction aims and climate-change mitigation actions (Schleussner et al. 2016). Forests play a pivotal role in these considerations as a potential C-sink that is manageable (Pan et al. 2011). To make informed decisions, knowledge on status and trends of the C-dynamic within forests are required. Countries that signed the Paris agreement and its precursors, therefore conduct annual greenhouse gas inventories (GHGI). Where available, National Forest Inventories (NFIs), are typically used to report C-stock gains and losses within forests and other wooded lands (McRoberts et al. 2020).

NFIs are sample surveys with a field component that (typically) consists of a net of systematically distributed sample plots (Tomppo et al. 2010) and are often the main source of information in the LULUCF (Land use, land-use change and forestry) part of the GHGI's AFOLU (agriculture, forestry, and other land use) sector. One of the largest contributors to the uncertainty of estimates within LULUCF are C-stock losses on forest land. The reason for this rather large uncertainty is that most of the C-stock losses result from final fellings that are rare events for NFIs because they are concentrated on fairly small areas with large C-stock losses. As a consequence, the number of sample plots with observed final fellings is small which in turn results in a large relative uncertainty (Strîmbu et al. 2017). Improving NFI-based estimates with remotely-sensed auxiliary data is therefore a natural choice (Mandallaz 2008, Ch. 6.2).

The strong correlation of airborne laser scanning (ALS) with forest attributes of importance for GHGIs is well established in the literature (Goetz and Dubayah 2011, Wulder et al. 2012, Maltamo et al. 2014). This correlation has been utilized to improve estimates of biomass stocks over large regions, typically using NFI sample plots to train models. Using ALS strip sampling, Gregoire et al. (2011) and Ståhl et al. (2011) have to this end developed model-assisted and model-based (hybrid) estimators that were employed within a county in Norway (Næsset et al. 2013). Ene et al. (2013) and Ringvall et al. (2016) proposed model-assisted and model-based estimators and applied them to post-strata consisting of administrative units. Extensions of these estimators to include full-coverage auxiliary data from optical satellites such as Landsat have been established using the hierarchical model-based framework (Saarela et al. 2015, Saarela et al. 2020) and Bayesian methods (Babcock et al. 2018). Wall-to-wall ALS, InSAR, and optical satellite data were compared for biomass and forest area estimation in a case study in Tanzania (Næsset et al. 2016). Considerable efficiency gains have been reported for national biomass estimates using wall-to-wall data sets of ALS data in Denmark (Magnussen et al. 2018, Magnussen and Nord-Larsen 2019). Also the utility of image-matching data (Breidenbach et al. 2016, Pulkkinen et al. 2018) or space-borne lidar (e.g., GEDI) in combination with Landsat (Saarela et al. 2018) and InSAR (Qi et al. 2019) satellite data to improve biomass estimates has been analyzed.

Whether with or without leveraging estimates with remotely-sensed data, C-stock changes can either be estimated directly using the observed change, for example at permanent sample plots, or indirectly by differencing stock estimates for two points in time. Estimating changes indirectly results in a net change estimate, which is either a net gain or net loss of stock. Estimating changes directly provides concurrent gross gain and gross loss estimates, which increases the transparency of the reporting which is an important concept in GHGIs (IPCC 2006, Vol. 1, Ch. 1.4) that facilitates the quality control of an inventory by expert reviewers. For smaller study areas, biomass change has been estimated directly and indirectly using repeated wall-to-wall ALS surveys (Skowronski et al. 2014, McRoberts et al. 2015, McRoberts et al. 2018, Esteban et al. 2019). Using repeated ALS strip sampling and NFI data, Ene et al. (2017) indirectly estimated biomass stock changes in Miombo woodlands of Tanzania. Using similar data from Norway, Strîmbu et al. (2017) compared model-assisted and model-based hybrid estimation for indirectly estimating biomass stock changes. Considerable efficiency gains compared to estimates based on field data only were reported.

For countries that do not have an established NFI, synthetic estimators are sometimes the only option to estimate biomass stocks or changes which makes a rigorous uncertainty estimation a complex task (McRoberts et al. 2020). In our context, synthetic estimators typically refer to totalizing model predictions over larger areas, e.g. the sum over all pixels of a map, without the possibility to correct for systematic errors in the applied model. For examples, Csillik et al. (2019) used a combination of ALS and optical data from Planet's Dove satellites to map the C-stock and estimate stock changes in Peru and Solberg et al. (2018) used repeated space-borne InSAR data to map and estimate the C stock-change of Uganda. Models were in both cases fit using field data located outside the study areas. Landsat, space-borne lidar (i.e., GLAS), various map products, and a small number of sample plots were used to map and estimate the total tree biomass in Kenya (Rodríguez-Veiga et al. 2020).

A relevant data set with respect to areas of C-stock losses is Global Forest Change; a Landsat-based service that provides an annual Forest Cover Loss (FCL) map with an approximately 30 m pixel size on a global scale (Hansen et al. 2013). In countries with small deforestation rates, FCL mostly provides annually updated information on harvests (i.e., final fellings). This was utilized by Rossi et al. (2019) to estimate the harvested area in a Norwegian mountain catchment based on post-stratification. They found that FCL detected 94% of all harvest sites but underestimated harvest area by more than 50% before 2012. Presumably due to a new Landsat satellite, FCL estimated harvest area reliably after 2012. In combination with ALS data that provide information about the canopy height before a change, FCL can be used to map biomass or timber volume losses after 2012 (NIBIO 2020). Breidenbach et al. (2020) summarized the harvest volume maps for municipalities in south-eastern Norway and found that these synthetic estimates, despite for an unknown bias of the synthetic estimator, were well correlated to official harvest statistics.

Ceccherini et al. (2020) used FCL for synthetic estimates (i.e. summation of map pixels) of harvest area and its temporal development from including the related C-stock changes in the EU-countries. They report harvest area increases of more than 30 and 50% when comparing the reference period of 2004-2015 with 2016-2018 in Sweden and Finland, respectively. However, their estimates did not include an uncertainty margin and diverged considerably from official statistics. Furthermore, their reference period included the years where Rossi et al. (2019) found that FCL underestimated harvest areas by 50% in Norway. A part of the reported increase of harvest areas by Ceccherini et al. (2020) may therefore be attributed to changes in the Landsat sensor or the model underlying FCL and not to actual harvest area increases. While synthetic estimators used by Ceccherini et al. (2020) are generally biased, synthetic estimates can be accurate and the bias negligible if the underlying model fits as it was found by Breidenbach et al. (2020) for synthetic estimates of harvest volume in 2016. Uncertainties from the ignored and possibly substantial model lack-of-fit will, however, remain unknown. Model-assisted and other design-based estimators (McRoberts et al. 2016) can mitigate the potential bias inherent in maps using additional reference observations.

As described above, a number of studies have shown the potential of remotely-sensed data to improve estimates of variables required in GHGIs compared to using only field-based data. However, the methods were often not targeted on GHGIs where annual estimates of gains and losses are required (IPCC 2006, Vol. 4, Annex 1). The aim of this study was therefore to assemble a model-assisted procedure to improve annual NFI-based estimates relevant for GHGIs for specific land-use categories within the AFOLU sector using auxiliary data. In case studies covering (large parts of) Norway, Latvia, Denmark, and Sweden, model-assisted (MA) estimates were compared to the basic expansion (BE) estimates using only field data. While the variable of interest was gross C-stock losses as the main contributor of uncertainty in the living biomass pool, the described methods are also applicable for estimates of C-stocks or gains thereof. Auxiliary data were FCL and, in regions where they were available, ALS data in combination with FCL. While most studies on biomass change used repeated ALS acquisitions, ALS data from one point in time were used here because repeated ALS campaigns are still rare for large regions. In this application, ALS data describe the forest structure in terms of C-stock before a potential loss of forest canopy – information which is missing in FCL. While most studies based on NFI data used the pooled sample, i.e. sample plots acquired over several years, we show the advantages of using annual samples.

We first define the estimators in a general form before we describe how we used them with the available remotely-sensed and NFI data. After a comparative overview of the main findings across countries, results are presented for each case study and are then collectively discussed.

## 3 Methods

### 3.1 Overview

Many NFIs utilize an interpenetrating panel design where a certain proportion of all sample plots is evenly distributed over the survey region and measured every year (Vidal et al. 2016). For example, all the NFIs considered here, use interpenetrating panels that consist of 1/5$^{th}$ of all plots that are systematically distributed over the whole country or stratum. Under such designs, the two options for estimating parameters of the sampled population are i) utilizing the *pooled* sample consisting of all panels of plots, or ii) averaging over estimates based on the *annual* sample consisting of the panel of plots measured in one year (Heikkinen et al. 2012). For estimators based on the sample data only – we will refer to those as *basic expansion* (BE) estimators – there is no difference between the two approaches. However, for *model-assisted* (MA) estimators that capitalize on remotely sensed data to improve precision, the difference in estimated *variances* can be large if the utilized data and models change over time.

Many NFIs use stratification to sample more intensively within highly productive regions than in less productive or less accessible regions. In NFIs, the observations are made on *sample plots* which often are clusters consisting of *sub-plots*. We denote $y_i$ as the mean over the variable of interest observed at the sub-plots of the i-th sample plot. To estimate the population parameter of interest for a certain domain such as forest, grassland, or any other land-use category required, a domain indicator variable $I_d$ is used. This indicator is 1 if the sample plot belongs to the domain of interest and 0 otherwise, when calculating the mean

$$y_i = \frac{\sum_j^{m_i} I_d y_{ij}}{m_i} \quad (1)$$

where $y_{ij}$ is the observed value of the variable of interest at the j-th sub-plot of the i-th sample plot (cluster) (Mandallaz 2008, p. 65). The number of sub-plots $m_i$ is independent of $I_d$ and typically fixed within a stratum but can vary due to the irregular shape of the country or stratum. That is, $m_i$ is the number of plots on land, which usually is constant but can vary for clusters located close to the coast or along country boarders. In NFIs that do not use clustered sample plots, $m_i$ is always one and $y_i = y_{ij}$ in all cases.

In the next sections we introduce the notation for the BE and MA estimators based on pooled and annual data before we describe how we utilized them with our specific NFI and remotely-sensed data.

### 3.2   Basic expansion estimators using only field data

### 3.2.1   Estimators based on the pooled sample

BE (basic expansion) estimators use only the sample plot information without leveraging remotely-sensed auxiliary information and can be used to directly or indirectly estimate C-stock changes.

Based on the pooled sample $s_P$ with $n_P$ sample plots, the total of a population parameter of interest is estimated by multiplying the estimated mean of the population parameter of interest $\hat{Y}^{BE,P}$ with the (known) area of the country or stratum $\lambda$

$$\hat{t}^{BE,P} = \lambda \hat{Y}^{BE,P} = \lambda \frac{\sum_{i \in s_P} m_i y_i}{\sum_{i \in s_P} m_i} \quad (2)$$

where the superscripts identify the BE estimator as being based on the pooled sample. $\hat{Y}^{BE,P}$ is the mean over all sub-plots ignoring the cluster structure (Mandallaz 2008, p. 66) and is the ratio of two random variables because $m_i$ is not fixed. Therefore, its variance is estimated as

$$\hat{V}(\hat{Y}^{BE,P}) = \frac{1}{n_P} \sum_{i \in s_P} \left(\frac{m_i}{\bar{m}}\right)^2 s^2 \quad (3)$$

where $s^2 = \frac{(y_i - \hat{Y}^{BE,P})^2}{(n_P - 1)}$ is the sample variance and $\bar{m} = \frac{1}{n_P} \sum_{i \in s_P} m_i$ is the average number of sub-plots (Mandallaz 2008, p. 68). The variance of the total is estimated by multiplying the squared area of the country or stratum with the variance of the mean

$$\hat{V}(\hat{t}^{BE,P}) = \lambda^2 \hat{V}(\hat{Y}^{BE,P}). \quad (4)$$

The standard error is the square root of the estimated variance $SE(\cdot) = \sqrt{\hat{V}(\cdot)}$. For NFIs that do not use clustered plots (m=1 in all cases), estimators (2) and (3) reduce to $\hat{t}^{BE,P} = \lambda \hat{Y}^{BE,P} = \lambda \frac{1}{n_P} \sum_{i \in s_P} y_i$ and $\hat{V}(\hat{Y}^{BE,P}) = \frac{1}{n_P} \sum_{i \in s_P} s^2$, respectively.

In the case of stratified sampling, which is very common in NFIs, the totals and variances are estimated separately for each stratum and then added. Because this is the case for all described estimators, we therefore do not describe the estimators with an index for strata for improved readability.

We assume random sampling and accept that the variances are likely conservative due to systematic sampling in NFIs; for a discussion on other options we refer to Magnussen et al. (2020) and Räty et al. (2020).

### 3.2.2 Estimators based on the annual panel

Estimators for annual panels are obtained by substituting the pooled sample $s_P$ with the annual sample $s_t$ and the superscript P with a in the estimators and symbols of Section 3.2.1. For example, the total based on the panel of plots sampled in year $t = \{1, \dots, T\}$ is

$$\hat{t}_t^{BE,a} = \lambda \hat{Y}_t^{BE,a}. \tag{5}$$

The average total estimate $\hat{t}^{BE,A}$ and its variance estimate $\hat{V}(\hat{t}^{BE,A})$ over all annual estimates are given by the weighted means

$$\hat{t}^{BE,A} = \frac{\sum_{t \in T} n_t \hat{t}^{BE,a}}{n_P} \tag{6}$$

where $n_t$ is the number of plots in the panel of year $t$, $n_P = \sum_{t \in T} n_t$ is the total number of sample plots and

$$\hat{V}(\hat{t}^{BE,A}) = \frac{\sum_{t \in T} n_t^2 \hat{V}(\hat{t}^{BE,a})}{n_P^2}. \tag{7}$$

In the case of BE estimators, the averaged total, mean, and variance estimators equal to the pooled variants and are given here just for a clearer description of the MA estimators below.

## 3.3 Model-assisted estimators

MA (model-assisted) estimation with the pooled or annual samples is analogous to the BE estimators. For improved readability, we therefore drop the superscripts $P$, $a$, and $A$ here. For MA estimation, a working model of arbitrary type is used that yields predictions of the variable of interest. We are interested in working models applicable at the sub-plot level that result in the predictions $\hat{y}_{ij}$. The MA estimator is given as a sum of the synthetic estimate $\hat{t}^S$ and an estimated correction factor $\hat{t}^C$

$$\hat{t}^{MA} = \hat{t}^S + \hat{t}^C. \tag{8}$$

If the working model is used for mapping the variable of interest based on some auxiliary information, for example extracted from remotely-sensed data, the synthetic estimate is the sum of all pixel values in the area of interest. In this case, the model is the mathematical device that translated the remotely-sensed information to the values in the mapped pixels. As the name suggests, the correction factor corrects for potential model lack-of-fit. The correction factor is estimated using the BE estimator (2) by substituting the observed value $y_i$ with the model residual $e_i$ in eq. (1)

$$e_i = \frac{\sum_j^{m_i} I_d(y_{ij} - \hat{y}_{ij})}{m_i} = \frac{\sum_j^{m_i} I_d e_{ij}}{m_i} \tag{9}$$

where $e_{ij}$ is the model residual at the sub-plot level. In analogy, the BE variance estimator (3) is used to estimate the variance of the MA estimate denoted by $\widehat{V}(\hat{t}^{MA})$ with $s^2 = \frac{\sum_{i \in s}(e_i - \bar{e})^2}{(n-1)}$ and the mean residual $\bar{e}$ (Cochran 1977, p. 195, McRoberts et al. 2014, p. 278). If the variance of the MA estimate is smaller than the variance of the BE estimate, the MA estimate will be more precise and efficient than the BE estimate.

The relative efficiency *RE* is the ratio of variances

$$\text{RE} = \widehat{V}(\hat{t}^{BE}) / \widehat{V}(\hat{t}^{MA}). \tag{10}$$

and can be used as a measure of efficiency of the MA estimate. A RE >1 means that the MA estimate is more efficient than the BE estimate. For example, RE=1.5 means an efficiency increase that would require 50% more field sample plots for the BE estimator to obtain the same precision as the MA estimator.

The pooled and average annual MA estimates are the same, if the NFI data and the applied auxiliary data and models are identical. However, as we will see in the case studies, it is an advantage of the annual estimates, that the applied auxiliary data and models can be different from the pooled estimate.

## 3.4 Application of the estimators

### 3.4.1 Generalities

The methods described here are focused on continuous NFIs utilizing interpenetrating panel designs that allow annual estimates for the whole country. Our variable of interest $y_{ij}$ is the annual carbon (C) stock loss (t/ha/year) observed at the permanent plot *i* and sub-plot *j*. In our case, annual means that the observed C-stock loss is divided by the number of years since the last measurement, which is five for the NFIs involved. Note that a C-stock loss can be due to human interventions such as harvests and deforestation, but also due to natural causes such as wild fires or storms, or the natural mortality of single trees. Some NFIs also record the likely year of a change in forest structure, but this information is not used here. We assume the C-stock loss to be measured without error. While C-stock loss is the variable of interest here, the estimators described above are generally applicable also for estimating other variables of interest such as C-stock gains or areas by land-use classes.

The population parameter to estimate is the total C-stock loss (t) which we will estimate over all land use categories by setting the indicator variable [eq. (1)] to $I_d=1$ in all cases. Estimates for forest land are obtained by setting the indicator variable to $I_d=1$ if the current land use is forest and $I_d=0$ otherwise.

Auxiliary data available globally with respect to C-stock loss are, for example, the Landsat-based Global Forest Change forest cover loss maps (FCL) (Hansen et al. 2013). While FCL provides the information in which year a forest canopy loss likely happened, it does not provide information on how much C was lost due to the change. However, if auxiliary data related to the C-stock *before* the change are available, then the C-stock loss can be predicted by assuming that the full C-stock was removed where FCL detected a canopy loss. We use airborne laser scanning (ALS) data for predicting the C-stock before the canopy loss.

### 3.4.2 Average annual estimates

In the NFIs considered, each permanent sample plot is revisited every 5$^{th}$ year which results in five annual panels that are the samples $s_t$. For an annual estimate based on the sample plots of one panel, say $t = 2018$, it is assumed that any C-stock loss occurred between 2014 and 2018. The FCL map is recoded such that pixels indicating canopy loss in 2014-2018 ([$t$-4] - $t$) are set to the value of 1 and all other pixels are set to the value of 0.

MA estimators using two different models are compared in the case studies. In one model, just information from FCL is utilized and the predictions are given by

$$\hat{y}_{ij} = \begin{cases} \bar{y}_{CL} & if\ FCL = 1 \\ \bar{y}_N & if\ FCL = 0 \end{cases} \quad (11)$$

where $\bar{y}_{CL}$ and $\bar{y}_N$ is the mean C-stock loss observed at subplots with and without FCL, respectively. This model will be referred to as the *FCL model* and the post-fix *FCL* is added to the estimators utilizing this model.

The other model utilizes ALS data in addition to FCL (*ALS-FCL model*) and the post-fix *ALS-FCL* is added to the estimators utilizing this model. Model predictions are given by

$$\hat{y}_{ij} = \begin{cases} \widehat{cs}_{ij}/5 & \text{if FCL} = 1 \text{ within ALS coverage} \\ \bar{y}_{CL} & \text{if FCL} = 1 \text{ outside ALS coverage} \\ \bar{y}_N & \text{if FCL} = 0 \end{cases} \quad (12)$$

where

$$\widehat{cs}_{ij} = \hat{\beta}_0 + \hat{\beta}_1 x_{ij} + \hat{\beta}_2 x_{ij}^2 \quad (13)$$

is the predicted C-stock that is assumed to be completely lost in areas with FCL, $x_{ij}$ is the ALS mean height of first-returns at a sub-plot, and the $\hat{\beta}$s are estimated model parameters. Note that growth since the ALS acquisition was not modelled here and that any kind of model including non-parametric models could be used. The model parameters can be estimated with any data or may be taken from the literature because the MA estimator corrects for potential model lack-of-fit. However, the smaller the random error of the model, the more efficient is the MA estimator. In model (12), the predicted stock is divided by five ($\widehat{cs}_{ij}/5$) to obtain a mean annual stock change estimate given the five-year remeasurement interval of the NFIs considered here.

As an example, when using the ALS-FCL model with panel $t = 2018$, the ALS data must have been acquired in 2013 or before (ALS-year $\leq [t - 5]$) to ensure that the ALS data describe the forest structure before the canopy loss observed by FCL. Because nation-wide ALS campaigns only have started in the recent years, the areas with the required ALS data may currently still be limited in some countries.

The sample data and models are sufficient to estimate variances and REs, which are of most interest to determine if the suggested methodology is useful. For the point estimates (total C-stock loss), which are of main interest for users and stakeholders, also population-level synthetic estimates are required. Because the point estimates are of less interest in this study, we describe the calculation of the synthetic estimates in the Appendix.

### 3.4.3 Pooled estimates

For a pooled estimate using all panels, the reference periods for the remotely-sensed data change accordingly. We denote the five panels of the pooled sample as $t_1 - t_5$, for example 2014-2018. The FCL map is recoded such that pixels indicating canopy loss in 2010-2018 ($[t_1\text{-}4] - t_5$) are set to 1 and all other pixels are set to 0. For the MA estimator utilizing the ALS-FCL model (12), the ALS data must have been acquired in 2009 or before (ALS-year $\leq [t_1 - 5]$). It is to be expected that the average annual estimates have a smaller variance than the pooled estimates because the temporal mismatch of the FCL maps is bigger in the pooled estimate than in the annual estimates.

## 4 Case studies

### 4.1 Material in common for all countries

NFI data from Norway (NNFI), Sweden (SNFI), Denmark (DNFI), and Latvia (LNFI) are used. All considered NFIs have in common that they are based on interpenetrating panels that consist of 1/5[th] of the plots measured in one year. This allows annual estimates using the annual panels $t_1 - t_5$ = 2014-2018. While some of the NFIs also use temporary plots, we will only consider the permanent plots here where C-stock changes can be directly observed from repeated tree-level measurements. C-stock losses result from the removal of trees for which C (biomass) was predicted using models that utilize the measured dbh and other measured or predicted information such as tree heights. C-stock losses were summarized at the plot-level and scaled to annual per-ha values (t/ha/a). The total area covered by the case studies is larger than 70 Mha and recordings from more than 75,000 NFI (sub-) plots were used. In all NFIs, the gross C-stock loss was observed at the plots or sub-plots except in Sweden where the net C-stock loss was observed. While Denmark had a full ALS coverage from 2006, the other countries started later with their national campaigns and were only partially covered. An overview of the data used in the case studies is given in Table 1.

FCL in version 1.6 contains the mapped (predicted) years of canopy losses given Landsat timeseries (Hansen et al. 2013, UMD 2020). This dataset was used as auxiliary information in all countries. Point estimates were only calculated for the case of Norway as the emphasis of this study is on the utility of the proposed method, which can be judged based on the relative efficiency of the MA estimators.

Table 1: Overview of the data used in the case studies.

| Country | Study area | Area (Mha) | ALS coverage (%)* | Forest area | Sample grid | Clustering | # sample plots / sub-plots | Field-plot setup*** |
|---|---|---|---|---|---|---|---|---|
| Norway | Lowland stratum | 15.0 | 0 – 25 | 66% 9.9 Mha | 3×3 km | None, plot=sub-plot | 16,631 / no clustering | 250 m$^2$ full plot, 5 cm min dbh |
| Latvia | Whole country | 6.4 | 0 – 3 | 51% 3.3 Mha | 4×4 km | 4 sub-plots, 250x250 m square | 4,053 / 16,157 | 25-100-500 m$^2$ concentric, 2-6-14 cm dbh thresholds |
| Denmark | Whole country | 4.3 | 100 | 15% 0.6 Mha | 6×6 km** | 4 sub-plots, 200x200 m square | 3,706 / 14,437 | 38.48-314.16-706.86 m$^2$ concentric, 0-10-40 cm dbh thresholds |
| Sweden | Whole country | 45.1 | 3 – 32 | 62% 28.0 Mha | 11x11-26x26 km | 4-8 sub-plots, 300x300-1200x1200 m square | 4,036 / 28,547 | 3.14 - 38.48 - 314.16 m$^2$ concentric, 0-4-10 cm dbh thresholds |

* ALS coverages of relevance for the estimators used in this study. ** Permanent plots only. *** Only with respect to tree-level measurements of relevance for this study.

## 4.2 Comparison of the main results across countries

The use of FCL or ALS in combination with FCL reduced uncertainties of C-stock loss estimates over all land-use categories in all cases as REs ranged between 1.01 and 2.70 (Figure 1). While the largest improvements in terms of variance reduction of the BE estimator resulted from the use of FCL in the MA estimator, the additional use of ALS to describe the C-stock before a change in most cases further improved the estimates. The use of ALS, however, often improved the estimates only modestly and in some cases, it even increased the variance compared to just using FCL. Responsible for the latter were typically commission errors in FCL, which became pronounced when using ALS in addition. The efficiencies of MA estimators tended to increase with harvest activity, or more specifically, with the proportion of (sub-)plots with FCL (Figure 2). Consequently, countries with relatively small harvest areas like Denmark or Norway profited less from using remotely-sensed data than Sweden or Latvia.

Furthermore, the largest improvements in all countries were observed in 2018, when the harvest areas were relatively large. In that year, the MA estimates utilizing the ALS-FCL model were more precise than MA estimates utilizing the FCL model and, in comparison to the BE estimator, the SEs dropped by 2.4, 3.1, 2.1, and 2.6% in Norway, Sweden, Denmark, and Latvia, respectively.

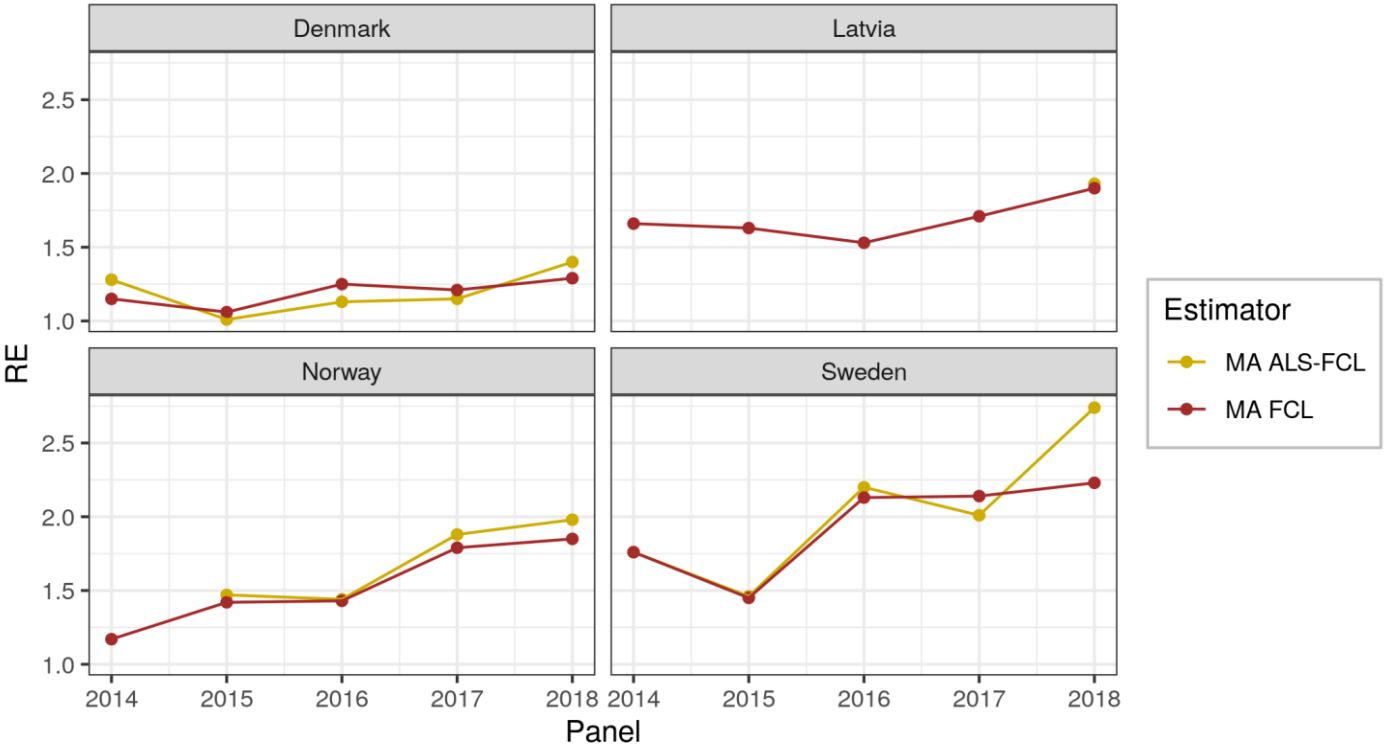

*Figure 1: Relative efficiency (RE) of annual C-stock loss estimates over all land-use categories per country for the MA estimators using the FCL model or the ALS-FCL model.*

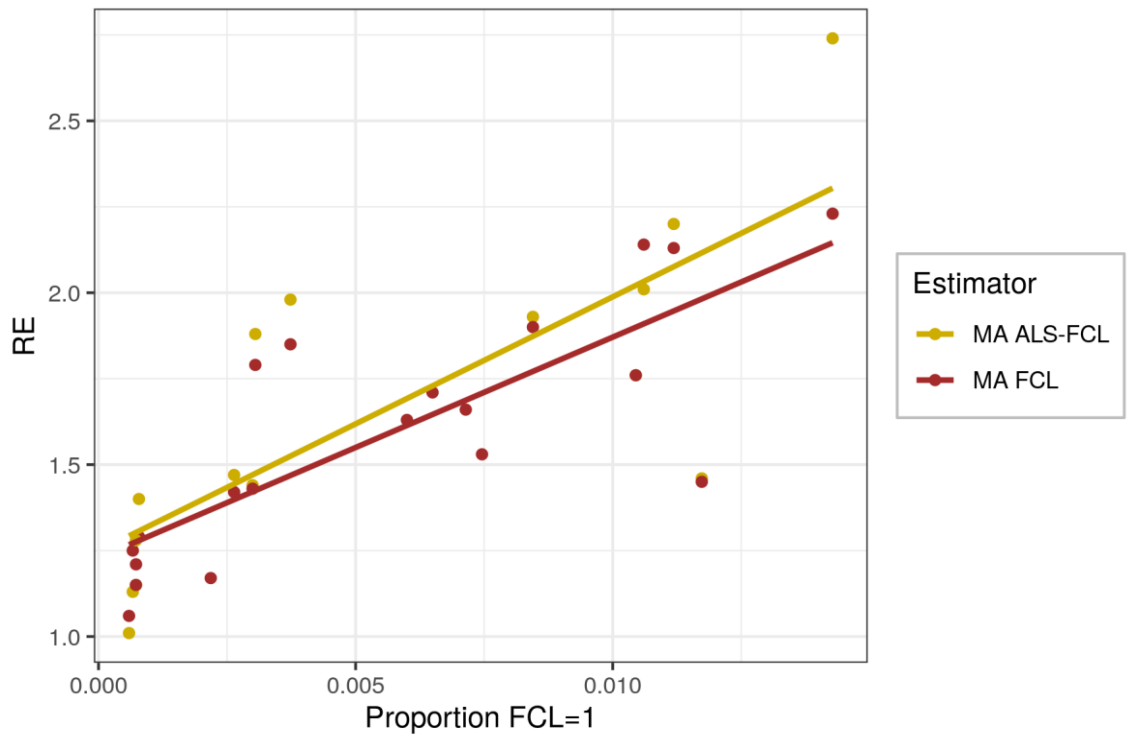

*Figure 2: Relative efficiency (RE) of annual C-stock loss estimates over all land-use categories from all countries and years for the MA estimators using the FCL model or the ALS-FCL model given the proportion of (sub-)plots with FCL=1.*

## 4.3   Norway

### 4.3.1   Material

The NNFI dataset consisted of 16,631 sample plots in the lowland stratum that are located on a 3×3 km grid. The lowland stratum covers approximately 46% of Norway's land area, more than 80% of the forest area, and more than 90% of the woody biomass (Breidenbach et al. 2020). The predominantly boreal forests in the study area are dominated by Norway spruce, Scots pine, and birch species (Breidenbach et al. 2020b).

As opposed to the other NFIs considered, the NNFI uses a single sample plot (i.e., not a cluster of sub-plots) for acquiring the main dendrometric variables such as timber volume, biomass, or C (Table 1). Tree heights are measured for a subsample with an expected number of 10 trees and the remaining tree heights are inferred based on these measurements and the measured dbh's. Above and below ground tree biomass is predicted using biomass models. More information on the applied procedures and models in the NNFI is available from Breidenbach et al. (2020).

The use of ALS data in forest management inventories has a long history in Norway that dates back to the mid-1990s (Næsset 2014). Since 2010, ALS data acquisitions are standardized and used by the Norwegian Mapping Authority in the ongoing development of a fine-resolution national digital terrain model (DTM) (Kartverket 2020). Since 2015, ALS data with a pulse density of either 2 or 5 per m$^2$ are specifically acquired for the national DTM and the ALS data available in the lowland NFI stratum were utilized in this study. Because the ALS data must have been acquired before a potential change in the forest canopy (ALS-year $\leq [t-5]$), no ALS data were available for the t=2014 panel. Therefore, the MA estimate using the ALS-FCL model for t=2014 and the pooled MA estimate using the ALS-FCL model were not possible because ALS data from 2009 or before would have been needed. For the 2015-2018 panels, the ALS coverage ranged from 10%-25%. The MA average annual estimate therefore combines the MA estimate for 2014 utilizing the FCL model with the MA estimates for 2015-2018 utilizing the ALS-FCL model. Because this is the combination of the estimates with the smallest variance, this combination will be referred to as the BEST-estimators.

### 4.3.2 Estimates

The parameters of the C-stock model (13) were estimated from the pooled 2014-2018 panels for which ALS data were available after excluding obviously outlying plots that likely resulted from harvests or other major changes in the time between ALS acquisition and field measurements. A total of 10,261 sample plots were used for fitting the model. Note that the subjective removal of outlying observations is unproblematic because the C-stock model is just a part of the working model whose possible systematic lack-of-fit is corrected for by the MA estimator. Because the ALS timing restriction required for estimation does not apply to the data used in the model, the number of sample plots with ALS information for modelling is much bigger than the number of sample plots with ALS information for estimation (Table 3). The estimated parameters of the ALS-based C-stock model (13) were $\hat{\beta}_0 = 1.18, \hat{\beta}_1 = 8.57, \hat{\beta}_2 = 0.087$. The parameters of the annual FCL models, which are also among the parameters of the ALS-FCL model are given in Table 2.

Table 2: Number of sample plots and parameter estimates of the annual working model [eq. (11)] for all land use categories. $\bar{y}_N$ is the estimated average C-stock loss at plots with FCL=0, $\bar{y}_{CL}$ is the estimated average C-stock loss at plots with FCL=1.

| Panel | $n_t$ per year | Parameter | Estimate (t/ha/a) | $n_t$ by FCL | $n_t$ covered by ALS |
|---|---|---|---|---|---|
| 2018 | 3351 | $\bar{y}_N$ | 0.32 | 3294 | 649 |
|  |  | $\bar{y}_{CL}$ | 17.15 | 57 | 16 |
| 2017 | 3314 | $\bar{y}_N$ | 0.3 | 3268 | 533 |
|  |  | $\bar{y}_{CL}$ | 16.82 | 46 | 7 |
| 2016 | 3329 | $\bar{y}_N$ | 0.37 | 3284 | 397 |
|  |  | $\bar{y}_{CL}$ | 14.76 | 45 | 4 |
| 2015 | 3262 | $\bar{y}_N$ | 0.35 | 3222 | 315 |
|  |  | $\bar{y}_{CL}$ | 14.09 | 40 | 7 |
| 2014 | 3375 | $\bar{y}_N$ | 0.36 | 3342 | 0 |
|  |  | $\bar{y}_{CL}$ | 8.9 | 33 | 0 |

A slightly increasing harvest area reported by FCL is visible in the considered period 2014-2018 and for 33-57 sample plots of the annual panels, FCL was observed ($FCL = 1$). Four to 16 of those plots were located within ALS coverage (Table 2). The average observed C-stock losses within areas with FCL =1 were considerably greater than in areas with FCL=0 (Figure 3A), suggesting that FCL detects harvests reasonably well. Although quite a few sample plots with C-stock loss were not detected by FCL (green dots with C-stock loss >0 in Figure 3A), no C-stock loss was observed for the vast majority of the plots without FCL. Due to their relatively small number, these omission errors of FCL had only a small influence, resulting in a median C-stock loss close to 0 (green horizontal line in Figure 3A). Commission errors (FCL=1 but no C-stock loss) were even less frequent than omission errors. The yellow points in Figure 3B follow the bisecting line reasonably well which suggests that the ALS-based C-stock model (13) may be useful to predict the C-stock loss in areas with FCL=1.

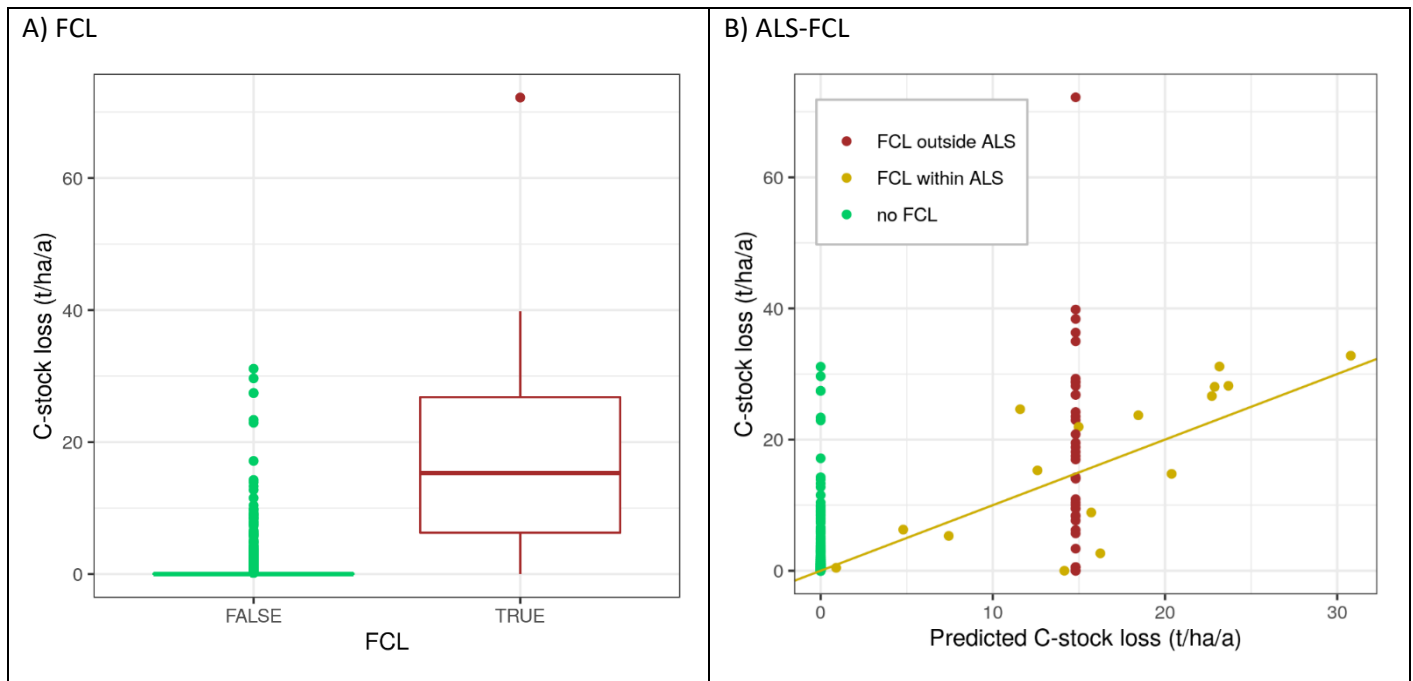

*Figure 3: Data of the 2018 panel of the Norwegian NFI used for fitting the parameters of the FCL and ALS-FCL models. A) Observed annual C-stock loss given forest-cover loss (FCL). B) Observed vs. predicted annual C-stock loss given forest-cover loss (FCL) or not.*

The BE total annual C-stock loss estimates that do not utilize remotely sensed data over all land-use classes for the years 2014-2018 ranged between 6.6 and 9.2 Mt (Table 3). For any given year, the MA estimates were not significantly different from the BE estimates (the 95% confidence intervals of all estimates include the point estimates of any other estimate in a given year). In the years for which ALS data were available (2015-2018), the REs of annual MA estimates using the ALS-FCL model ranged between 1.44 and 1.98. Furthermore, the MA estimates using the ALS-FCL model were always more efficient than the MA estimates using the FCL model. Within areas of ALS coverage, the MA estimator utilizing the ALS-FCL model facilitated REs up to 3 (see RE in the right part of Table 3).

The efficiency of the MA estimates increased from 2014 to 2018. For 2018, the year with the largest C-stock loss estimates and where the ALS coverage was already 25%, the SE drops from more than 9% for the BE estimates to less than 7% for estimates utilizing remotely sensed data in 2018 (Table 3). For forest land, the efficiency of the MA estimates reduced slightly compared to all land uses. This was because 95% of the C-stock losses stems from forest lands and the FCL and ALS-FCL models were the same for all land-use categories and forest land. However, some of the observed C-stock losses dropped to zero as they happened outside forest [see eq. (1)], which in most cases increased the residuals and hence the variance.

Table 3: Annual C-stock change estimates and annual C-stock change estimates by ALS coverage type (0=no ALS coverage, 1=ALS coverage). ALS-FCL indicates that the model utilizing ALS and FCL data was used whereas FCL indicates that the model utilizing only FCL data was used.

| Panel | $\hat{t}_t^{BE,a}$ ($10^6$ t) | SE($\hat{t}_t^{BE,a}$) (%) | $\hat{t}_t^{MA,a}$ ALS-FCL ($10^6$ t) | SE($\hat{t}_t^{MA,a}$) ALS-FCL (%) | RE ALS-FCL | $\hat{t}_t^{MA,a}$ FCL ($10^6$ t) | SE($\hat{t}_t^{MA,a}$) FCL (%) | RE FCL | ALS coverage type | RE ALS-FCL | RE FCL |
|---|---|---|---|---|---|---|---|---|---|---|---|
| All land use categories | | | | | | | | | | | |
| 2018 | 9.06 | 9.17 | 9.05 | 6.53 | 1.98 | 9.17 | 6.66 | 1.85 | 0 | 1.78 | 1.78 |
|  |  |  |  |  |  |  |  |  | 1 | 3.11 | 2.14 |
| 2017 | 7.91 | 9.57 | 8.36 | 6.60 | 1.88 | 8.32 | 6.80 | 1.79 | 0 | 1.88 | 1.88 |
|  |  |  |  |  |  |  |  |  | 1 | 1.88 | 1.25 |
| 2016 | 8.42 | 9.38 | 8.45 | 7.80 | 1.44 | 8.42 | 7.86 | 1.43 | 0 | 1.44 | 1.44 |
|  |  |  |  |  |  |  |  |  | 1 | 1.41 | 1.26 |
| 2015 | 7.71 | 9.44 | 7.48 | 8.04 | 1.47 | 7.55 | 8.08 | 1.42 | 0 | 1.38 | 1.38 |
|  |  |  |  |  |  |  |  |  | 1 | 1.94 | 1.61 |
| 2014 | 6.59 | 8.70 | - | - | - | 6.82 | 7.78 | 1.17 | 0 | 1.17 | 1.17 |
|  |  |  |  |  |  |  |  |  | 1 | - | - |
| Forest land | | | | | | | | | | | |
| 2018 | 8.53 | 9.43 | 8.52 | 7.12 | 1.76 | 8.64 | 7.19 | 1.68 | 0 | 1.64 | 1.64 |
|  |  |  |  |  |  |  |  |  | 1 | 2.38 | 1.85 |
| 2017 | 7.80 | 9.70 | 8.25 | 6.69 | 1.88 | 8.20 | 6.89 | 1.79 | 0 | 1.88 | 1.88 |
|  |  |  |  |  |  |  |  |  | 1 | 1.90 | 1.25 |
| 2016 | 8.13 | 9.59 | 8.16 | 8.14 | 1.38 | 8.13 | 8.21 | 1.37 | 0 | 1.38 | 1.38 |
|  |  |  |  |  |  |  |  |  | 1 | 1.41 | 1.26 |
| 2015 | 7.41 | 9.74 | 7.18 | 8.34 | 1.45 | 7.25 | 8.38 | 1.41 | 0 | 1.37 | 1.37 |
|  |  |  |  |  |  |  |  |  | 1 | 1.96 | 1.62 |
| 2014 | 6.31 | 8.98 | - | - | - | 6.54 | 8.07 | 1.15 | 0 | 1.15 | 1.15 |
|  |  |  |  |  |  |  |  |  | 1 | - | - |

Average annual estimates offer the possibility to combine the best estimates of each year. Because no ALS data were available for the estimate in 2014, we averaged the MA estimates utilizing the FCL model for 2014 with the MA estimates utilizing the ALS-FCL model for 2015-2018 (indicated by BEST in Table 4). This best average MA estimate was more efficient than the average over the MA estimates utilizing the FCL model and had an RE of 1.59.

The average annual estimate was considerably more efficient than the pooled MA estimate utilizing the FCL model. Pooled MA estimates utilizing ALS and FCL data were not possible because ALS data were not available before 2014. Due to the still increasing area of ALS coverage in Norway, the efficiency of MA estimates utilizing the ALS-FCL model can be assumed to further increase in the next years. Although the average annual estimates allow substantial variance reductions, the SEs only drop from around 4% for the BE estimate to around 3% for the average MA estimates (Table 4).

*Table 4: C-stock change estimates for the period 2014-2018. FCL indicates that the model utilizing FCL data was used whereas BEST indicates that the best annual estimate per year was used in the average annual estimate.*

| Land use category | Estimator | $\hat{t}_t^{BE}$ ($10^6$ t) | SE($\hat{t}_t^{BE}$) (%) | $\hat{t}_t^{MA}$ BEST ($10^6$ t) | SE($\hat{t}_t^{MA}$) BEST (%) | RE BEST | $\hat{t}_t^{MA}$ FCL ($10^6$ t) | SE($\hat{t}_t^{MA}$) FCL (%) | RE FCL |
|---|---|---|---|---|---|---|---|---|---|
| All | Average annual (A) | 7.94 | 4.17 | 8.03 | 3.27 | 1.59 | 8.06 | 3.33 | 1.52 |
| | Pooled (P) | 7.94 | 4.17 | - | - | - | 7.85 | 3.81 | 1.23 |
| Forest | Average annual (A) | 7.63 | 4.28 | 7.73 | 3.42 | 1.52 | 7.75 | 3.47 | 1.47 |
| | Pooled (P) | 7.63 | 4.28 | - | - | - | 7.55 | 3.93 | 1.21 |

Because of the clearly lower precision of the pooled estimator, we just present the results for the average annual estimators for the other case studies.

### 4.4 Sweden

The SNFI dataset consisted of 4,036 permanent sample plots (clusters) with a total of 28,547 sub-plots. Details about the SNFI can be found in Fridman et al. (2014) while some general information is available from Table 1. Sweden is predominantly covered by boreal forests dominated by Norway spruce, Scots pine, and birch species. The SNFI design is based on five geographical strata that are surveyed with decreasing sampling intensity towards the north. Permanent plots are located on regular grids with sizes ranging from 11x11 km in the south to 26x26 km in the northwestern parts of the country. There are 8 sub-plots per cluster except for one stratum with 4 sub-plots per cluster.

An ALS dataset covering almost all of Sweden was collected by the National Mapping Agency (Lantmäteriet) with the aim to create a new national digital terrain model (DTM). The ALS campaign started in 2009 and ended in 2019, with 97.5% of the productive forest land scanned by the end of 2015. The pulse density was 0.5-1.0 per m$^2$ and given the requirements of the estimators, ALS data from 2009-2013 were used here. As opposed to the other case-studies, the mean height of all ALS returns (instead of first returns) was available as the predictor variable and the estimated C-stock model parameters (13) were $\hat{\beta}_0 = -69.21, \hat{\beta}_1 = 28.67, \hat{\beta}_2 = 0.07$. Negative predictions were set to 0. The parameters of the annual FCL models, which are also among the parameters of the ALS-FCL model, are given in Table 5. The number of sub-plots with FCL=1 ranged from 158 in 2014 to 218 in 2018. Of those, 2 to 94 sub-plots were covered by ALS data that were acquired at least 5 years before the field measurement (Table 5). The distribution of C-stock losses given FCL is given in Figure 4A. As can be seen from Figure 4B, there were a number of sub-plots where the model predicted a C-stock loss although there was no loss according to the field observations. By checking aerial images, we found that this commission error, in most cases, was caused by FCL covering a larger area than the actual harvest site and the sample plot was well outside the harvest site.

Table 5: Number of sample plots and parameter estimates of the annual working model [eq. (11)] for all land use categories. $\bar{y}_N$ is the estimated average C-stock loss at plots with FCL=0, $\bar{y}_{CL}$ is the estimated average C-stock loss at plots with FCL=1.

| Panel | # clusters ($n_t$) | Parameter | Estimate (t/ha/a) | # sub-plots | # sub-plots covered by ALS* |
|---|---|---|---|---|---|
| 2018 | 825 | $\bar{y}_N$ | 0.92 | 5608 | - |
|  |  | $\bar{y}_{CL}$ | 39.46 | 218 | 94 |
| 2017 | 805 | $\bar{y}_N$ | 0.93 | 5506 | - |
|  |  | $\bar{y}_{CL}$ | 34.79 | 160 | 57 |
| 2016 | 797 | $\bar{y}_N$ | 1.05 | 5459 | - |
|  |  | $\bar{y}_{CL}$ | 41.21 | 168 | 59 |
| 2015 | 814 | $\bar{y}_N$ | 1.31 | 5573 | - |
|  |  | $\bar{y}_{CL}$ | 33.68 | 178 | 30 |
| 2014 | 798 | $\bar{y}_N$ | 1.19 | 5519 | - |
|  |  | $\bar{y}_{CL}$ | 40.44 | 158 | 2 |

* The number of sub-plots with ALS coverage that are not covered by FCL is irrelevant in this case study and therefore not given to avoid confusion.

The BE total annual C-stock loss estimates over all land-use categories ranged between 72 and 95 Mt (Table 6). The MA estimators were found to perform better than the BE estimators for all years and REs ranged between 1.46 and 2.70. Compared to the BE estimator, the SEs dropped by up to 3% when utilizing the ALS-FCL model in the MA estimator. The MA estimator using the ALS-FCL model performed better than the MA estimator using the FCL model for all years except 2017. Because more than 90% of the biomass losses come from forests, the MA estimators had a similar efficiency for forest lands as for all land-use categories (Table 6).

In the annual estimates of 2018, REs for all land-use categories within the five NFI strata ranged between 1.3 and 3.1 for the MA estimator using the FCL model, and between 1.1 and 5.1 for the MA estimator using the ALS-FCL model. Furthermore, the SEs of MA estimators using the FCL and ALS-FCL models were up to 11%-points and 10%-points smaller than the SEs of the BE estimator.

The averaged MA estimates had REs around 2 and SEs dropped from close to 3.4% for BE estimates to around 2.4% for MA estimates (Table 7). Most efficient, with a RE of 2.1, was the combination of the best annual MA estimates by using the ALS-FCL model in all years except for 2017 where the FCL model was used (estimates identified by BEST in Table 7).

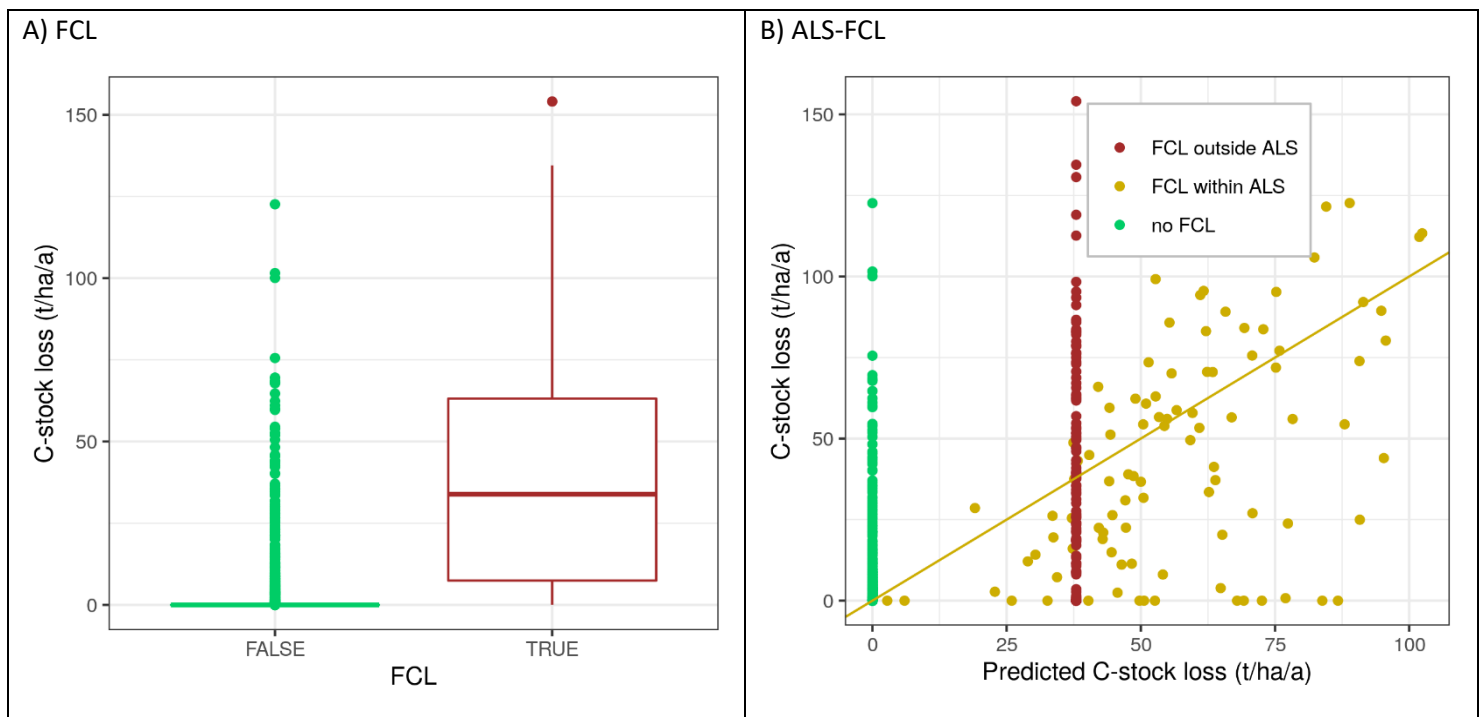

Figure 4: Data of the 2018 panel of the Swedish NFI used for fitting the parameters of the FCL and ALS-FCL models. A) Observed annual C-stock loss given forest-cover loss (FCL). B) Observed vs. predicted annual C-stock loss; predictions are based on the ALS-FCL model.

Table 6: Annual C-stock change estimates. ALS-FCL indicates that the model utilizing ALS and FCL data was used whereas FCL indicates that the model utilizing only FCL data was used.

| Panel | $\hat{t}_t^{BE}$ ($10^6$ t) | $SE(\hat{t}_t^{BE,a})$ (%) | $SE(\hat{t}_t^{MA,a})$ ALS-FCL (%) | RE ALS-FCL | $SE(\hat{t}_t^{MA,a})$ FCL (%) | RE FCL |
|---|---|---|---|---|---|---|
| All land-use categories | | | | | | |
| 2018 | 97.42 | 7.74 | 4.67 | 2.74 | 5.19 | 2.23 |
| 2017 | 71.78 | 7.43 | 5.24 | 2.01 | 5.08 | 2.14 |
| 2016 | 88.67 | 8.23 | 5.55 | 2.20 | 5.64 | 2.13 |
| 2015 | 93.48 | 7.14 | 5.91 | 1.46 | 5.93 | 1.45 |
| 2014 | 87.42 | 6.81 | 5.14 | 1.76 | 5.14 | 1.76 |
| Forest land | | | | | | |
| 2018 | 95.68 | 7.85 | 4.76 | 2.72 | 5.28 | 2.22 |
| 2017 | 67.39 | 7.67 | 5.33 | 2.07 | 5.17 | 2.21 |
| 2016 | 86.68 | 8.38 | 5.66 | 2.19 | 5.75 | 2.12 |
| 2015 | 90.44 | 7.36 | 6.11 | 1.45 | 6.12 | 1.45 |
| 2014 | 83.16 | 7.10 | 5.31 | 1.79 | 5.31 | 1.78 |

Table 7: Average annual C-stock change estimates for the period 2014-2018. ALS-FCL indicates that the model utilizing ALS and FCL data was used, FCL indicates that the model utilizing only FCL data was used, and BEST indicates that the best annual estimate per year was used.

| Land use category | $\hat{t}_t^{BE,A}$ ($10^6$ t) | SE($\hat{t}_t^{BE,A}$) (%) | SE($\hat{t}_t^{MA,A}$) ALS-FCL (%) | RE ALS-FCL | SE($\hat{t}_t^{MA,A}$) FCL (%) | RE FCL | SE($\hat{t}_t^{MA,A}$) BEST (%) | RE BEST |
|---|---|---|---|---|---|---|---|---|
| All | 87.82 | 3.37 | 2.39 | 2.00 | 2.42 | 1.95 | 2.38 | 2.01 |
| Forest land | 84.74 | 3.47 | 2.45 | 2.00 | 2.48 | 1.96 | 2.44 | 2.01 |

## 4.5 Denmark

The DNFI dataset consisted of 3,706 permanent sample plots with a total of 14,247 sub-plots. Details about the DNFI can be found in Nord-Larsen and Johannsen (2016) while some general information is available from Table 1. While we estimated the gross C-stock loss using the permanent sample plots, the official GHG reporting utilizes both these permanent and additional temporary sample plots for estimating a net C-stock loss or gain.

All of Denmark was ALS scanned for the first time in 2006 (Nord-Larsen and Schumacher 2012). These data were used here to predict the C-stock before a canopy cover loss indicated by FCL. ALS first-return mean heights were available for all sub-plots with forest cover and were assumed to be 0 for all other sub-plots. Because C-stock changes outside forest are rare and small in Denmark, we do not present results for all land use categories and forest land separately.

Between 9 and 12 subplots per panel had an FCL observation and no strong temporal trend was observed (Table 8). The estimated C-stock model parameters (13) were $\hat{\beta}_0 = 15.16, \hat{\beta}_1 = 10.95, \hat{\beta}_2 = -0.13$. The parameters of the annual FCL models, which are also among the parameters of the ALS-FCL model are given in Table 8.

Table 8: Number of sample plots and parameter estimates of the annual working model [eq. (11)] for all land use categories. $\bar{y}_N$ is the estimated average C-stock loss at plots with FCL=0, $\bar{y}_{CL}$ is the estimated average C-stock loss at plots with FCL=1.

| Panel | # clusters ($n_t$) | Parameter | Estimate (t/ha/a) | # sub-plots* |
|---|---|---|---|---|
| 2018 | 739 | $\bar{y}_N$ | 0.24 | 2833 |
|  |  | $\bar{y}_{CL}$ | 13.84 | 12 |
| 2017 | 752 | $\bar{y}_N$ | 0.22 | 2852 |
|  |  | $\bar{y}_{CL}$ | 11.64 | 11 |
| 2016 | 724 | $\bar{y}_N$ | 0.31 | 2789 |
|  |  | $\bar{y}_{CL}$ | 15.51 | 10 |
| 2015 | 760 | $\bar{y}_N$ | 0.28 | 2910 |
|  |  | $\bar{y}_{CL}$ | 7.86 | 9 |
| 2014 | 731 | $\bar{y}_N$ | 0.31 | 2810 |
|  |  | $\bar{y}_{CL}$ | 10.89 | 11 |

*All sub-plots with FCL have ALS coverage.

The BE total annual C-stock loss estimates over all land-use categories ranged between 1.12 and 1.56 Mt and the REs of annual MA estimators using the FCL model ranged between 1.01 and 1.40 (Table 9). Compared to the BE estimator, the SEs dropped by up to 2% when utilizing the ALS-FCL model in the MA estimator. The MA estimator using the ALS-FCL model was more efficient than the MA estimator using the FCL model in just two years (Table 9). In 2015, the MA estimators were only slightly more efficient than the BE estimator. In that year, the fewest number of sub-plots had an FCL coverage and a tendency of the ALS-FCL model to overpredict the observed C-stock loss was more pronounced than in other years (Figure 5A and B).

Table 9: Annual C-stock change estimates. ALS-FCL indicates that the model utilizing ALS and FCL data was used whereas FCL indicates that the model utilizing only FCL data was used.

| Panel | $\hat{t}_t^{BE}$ ($10^6$ t) | SE($\hat{t}_t^{BE,a}$) (%) | SE($\hat{t}_t^{MA,a}$) ALS-FCL (%) | RE ALS-FCL | SE($\hat{t}_t^{MA,a}$) FCL (%) | RE FCL |
|---|---|---|---|---|---|---|
| 2018 | 1.26 | 13.98 | 11.83 | 1.40 | 12.33 | 1.29 |
| 2017 | 1.12 | 13.64 | 12.70 | 1.15 | 12.39 | 1.21 |
| 2016 | 1.56 | 12.78 | 12.01 | 1.13 | 11.44 | 1.25 |
| 2015 | 1.31 | 12.14 | 12.11 | 1.01 | 11.75 | 1.06 |
| 2014 | 1.49 | 14.46 | 12.65 | 1.28 | 13.16 | 1.15 |

The averaged MA estimates were more efficient than the BE estimate (Table 10). Most efficient, with a RE of 1.24, was the combination of the best annual MA estimates by using the ALS-FCL model in 2018 and 2014, and the FCL model in the other years (estimates identified by BEST in Table 10).

Table 10: Average annual C-stock change estimates for the period 2014-2018 for all land-use categories. ALS-FCL indicates that the model utilizing ALS and FCL data was used, FCL indicates that the model utilizing only FCL data was used, and BEST indicates that the best annual estimate per year was used.

| $\hat{t}_t^{BE,A}$ ($10^6$ t) | SE($\hat{t}_t^{BE,A}$) (%) | SE($\hat{t}_t^{MA,A}$) ALS-FCL (%) | RE ALS-FCL | SE($\hat{t}_t^{MA,A}$) FCL (%) | RE FCL | SE($\hat{t}_t^{MA,A}$) BEST (%) | RE BEST |
|---|---|---|---|---|---|---|---|
| 1.35 | 6.03 | 5.51 | 1.19 | 5.47 | 1.20 | 5.40 | 1.24 |

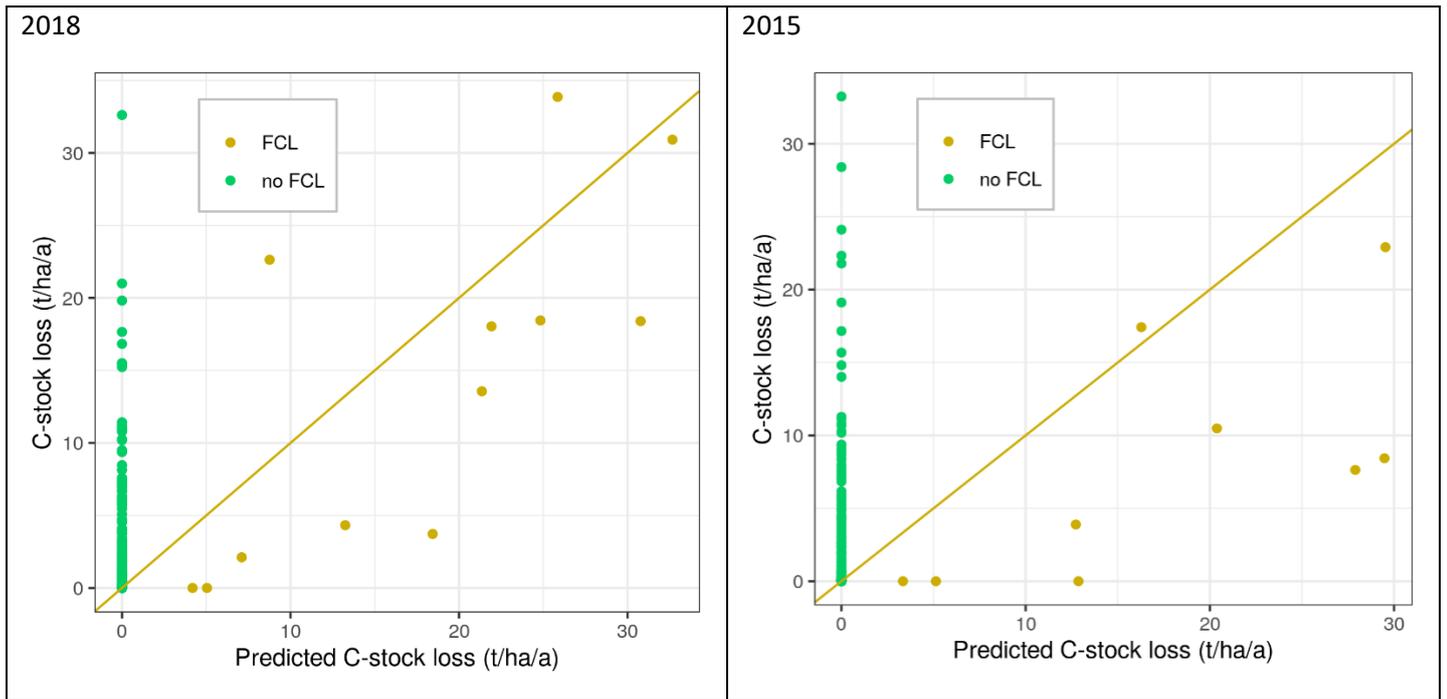

*Figure 5: Observed vs. predicted annual C-stock loss given forest-cover loss (FCL) or not at the sub-plots of the 2018 and 2015 panels of the Danish NFI.*

## 4.6 Latvia

The LNFI dataset consisted of 4,053 permanent sample plots with a total of 16,157 sub-plots. Details about the LNFI can be found in Jansons and Licite (2010) while some general information is available in Table 1. The hemi boreal forests in the study area largely are dominated by Scots pine, birch species, and Norway spruce. The national ALS campaign started in 2013 and covered approximately 3% of the land area in that year with a minimum pulse density of 4/m². Given our requirements, only these data were used to predict the C-stock before a canopy cover loss indicated by FCL for the annual estimate in 2018. Between 91 and 129 sub-plots of the annual panels had an FCL=1 and although no clear trend was observed, the number of sub-plots with FCL=1 was larger in the 2018 panel than in previous panels (Table 11). While the 3% ALS area coverage corresponded to 100 sub-plots of the 2018 panel with ALS data, only four had a cover loss observed by FCL (Figure 6). The estimated C-stock model parameters (13) were $\hat{\beta}_0 = -0.02, \hat{\beta}_1 = 8.25, \hat{\beta}_2 = 0.17$. The parameters of the annual FCL models, which are also among the parameters of the ALS-FCL model are given in Table 11.

Table 11: Number of sample plots and parameter estimates of the annual working model [eq. (11)] for all land use categories. $\bar{y}_N$ is the estimated average C-stock loss at plots with FCL=0, $\bar{y}_{CL}$ is the estimated average C-stock loss at plots with FCL=1.

| Panel | # clusters ($n_t$) | Parameter | Estimate (t/ha/a) | # sub-plots | # sub-plots with ALS coverage* |
|---|---|---|---|---|---|
| 2018 | 809 | $\bar{y}_N$ | 0.73 | 3119 | - |
|  |  | $\bar{y}_{CL}$ | 23.07 | 129 | 4 |
| 2017 | 811 | $\bar{y}_N$ | 0.79 | 3147 | - |
|  |  | $\bar{y}_{CL}$ | 23.91 | 98 | 0 |
| 2016 | 815 | $\bar{y}_N$ | 0.86 | 3151 | - |
|  |  | $\bar{y}_{CL}$ | 22.15 | 112 | 0 |
| 2015 | 810 | $\bar{y}_N$ | 0.88 | 3149 | - |
|  |  | $\bar{y}_{CL}$ | 23.27 | 91 | 0 |
| 2014 | 808 | $\bar{y}_N$ | 0.95 | 3140 | - |
|  |  | $\bar{y}_{CL}$ | 23.43 | 108 | 0 |

* The number of sub-plots with ALS coverage that are not covered by FCL is irrelevant in this case study and therefore not given to avoid confusion.

The BE total annual C-stock loss estimates over all land-use categories ranged between 9.38 and 10.45 Mt (Table 12). The REs of annual MA estimates using the FCL model ranged between 1.53 and 1.90. Although ALS covered only a relatively small area, the MA estimator using the ALS-FCL model was slightly more efficient than the MA estimator using the FCL model. The SEs dropped from above 8% for the BE estimates to around 6% for the MA estimates. Because more than 90% of the biomass losses come from forests, the MA estimators had a similar efficiency for forest lands as for all land-use categories (Table 12). The averaged MA estimates had REs around 1.70 and SEs dropped from close to 4% to around 3% for MA estimates (Table 13).

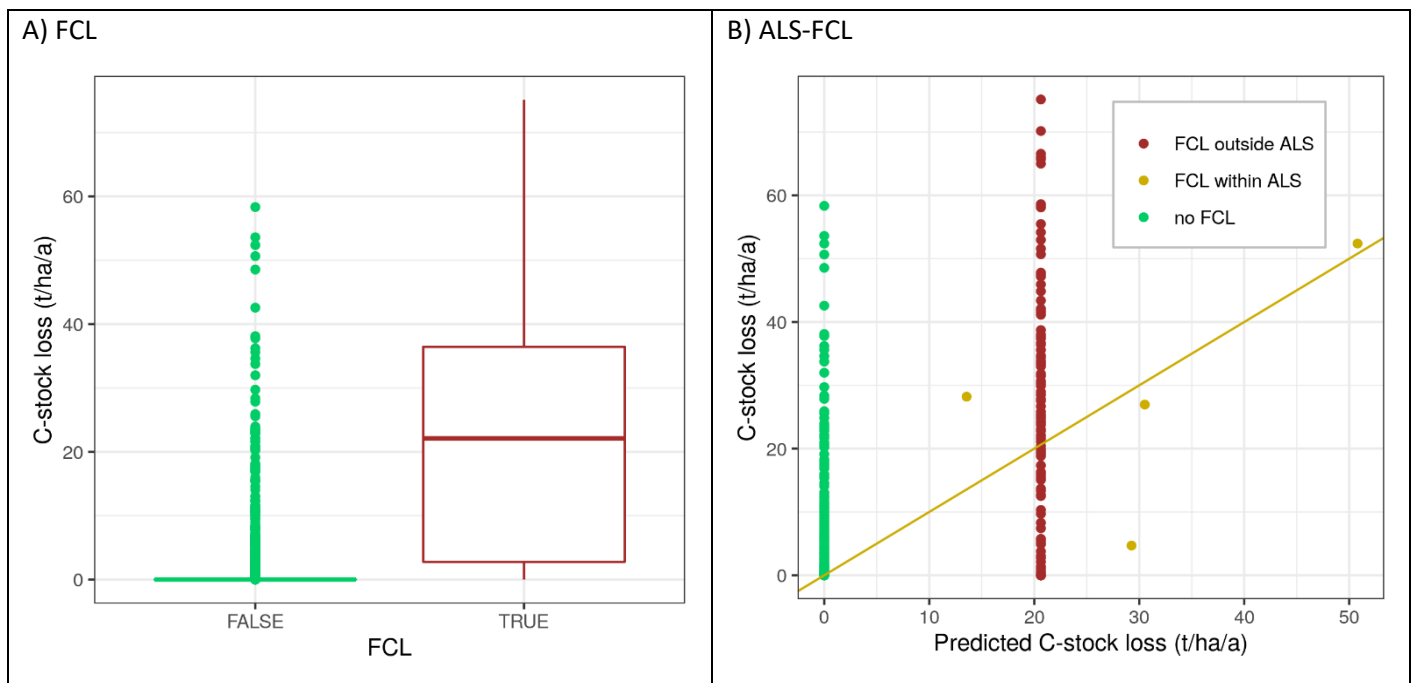

*Figure 6: Data of the 2018 panel of the Latvian NFI used for fitting the parameters of the FCL and ALS-FCL models. A) Observed annual C-stock loss given forest-cover loss (FCL). B) Observed vs. predicted annual C-stock loss; predictions are based on the ALS-FCL model.*

*Table 12: Annual C-stock change estimates. ALS-FCL indicates that the model utilizing ALS and FCL data was used whereas FCL indicates that the model utilizing only FCL data was used.*

| Panel | $\hat{t}_t^{BE}$ ($10^6$ t) | $SE(\hat{t}_t^{BE,a})$ (%) | $SE(\hat{t}_t^{MA,a})$ ALS-FCL (%) | RE ALS-FCL | $SE(\hat{t}_t^{MA,a})$ FCL (%) | RE FCL |
|---|---|---|---|---|---|---|
| All land-use categories | | | | | | |
| 2018 | 9.97 | 9.32 | 6.72 | 1.93 | 6.76 | 1.90 |
| 2017 | 9.44 | 8.54 | | | 6.53 | 1.71 |
| 2016 | 9.89 | 8.02 | | | 6.48 | 1.53 |
| 2015 | 9.37 | 8.02 | | | 6.28 | 1.63 |
| 2014 | 10.45 | 8.09 | | | 6.29 | 1.66 |
| Forest land | | | | | | |
| 2018 | 9.49 | 9.74 | 7.01 | 1.93 | 7.06 | 1.90 |
| 2017 | 8.80 | 9.09 | | | 6.92 | 1.73 |
| 2016 | 9.33 | 8.43 | | | 6.78 | 1.54 |
| 2015 | 8.84 | 8.40 | | | 6.61 | 1.62 |
| 2014 | 9.88 | 8.41 | | | 6.46 | 1.69 |

Table 13: Average annual C-stock change estimates for the period 2014-2018. FCL indicates that the model utilizing only FCL data was used and BEST indicates that the best annual estimate per year was used.

| Land use category | $\hat{t}_t^{BE,A}$ ($10^6$ t) | SE($\hat{t}_t^{BE,A}$) (%) | SE($\hat{t}_t^{MA,A}$) FCL (%) | RE FCL | SE($\hat{t}_t^{MA,A}$) BEST (%) | RE BEST |
|---|---|---|---|---|---|---|
| All | 9.83 | 3.77 | 2.89 | 1.69 | 2.89 | 1.70 |
| Forest land | 9.27 | 3.95 | 3.03 | 1.70 | 3.02 | 1.71 |

## 5 Discussion

In this study we found that a global dataset of forest cover losses (FCL) can improve estimates of C-losses in a design-based framework using NFI data. The additional use of one acquisition of ALS data in combination with FCL often further improved the estimates. The 3-dimensional information provided by ALS in this case predicts the expected C-stock before a canopy loss predicted by the 2-dimensional Landsat time-series that constitutes the basis of FCL (Hansen et al. 2013). In addition, the use of field-based NFI data in MA (model-assisted) estimators ensured that potential systematic model lack-of-fit was mitigated (Särndal et al. 1992). C-stock losses have been estimated using FCL and other satellite data only (Ceccherini et al. 2020). However, systematic errors that have been found in FCL especially before 2012 (Rossi et al. 2019) can then not be mitigated. This, in combination with the fact that uncertainties cannot be reliably estimated, may result in sub-optimal or even detrimental policy measures.

Most studies analyzing changes in biomass or C-stocks have so far used two ALS acquisitions, where one describes the status before, and the other the status after a change (McRoberts et al. 2015, Ene et al. 2017, Strîmbu et al. 2017). The advantage of using just one ALS acquisition for estimating C-stock losses as proposed here, is that repeated acquisitions over large areas still are rare due to the large costs involved. The clear advantage of using two ALS acquisitions is that not all canopy losses indicated by FCL are equivalent to the full removal of the C-stocks which is assumed in our approach. While the full removal of the C-stock through clear-felling is common in the boreal ecotone (Sweden and Norway), this is less the case in the hemi-boreal (Latvia) and temperate ecotone (Denmark) albeit also in the boreal ecotone remaining shelter wood, retention trees, and seed trees are common. In fact, the ALS-FCL model had a slight tendency to overpredict the C-stock change which was most visible in Sweden and Denmark. This could explain why the use of ALS data in addition to FCL often did not improve estimates in Denmark. Besides smaller harvest levels in Denmark than in the other countries, this may also be explained by the Danish ALS data that were acquired in 2006, while the ALS data in Norway and Sweden were acquired closer to the period under scrutiny. For Latvia, the area covered by ALS was yet too small to recognize a pattern in the model residuals.

Rossi et al. (2019) reported that FCL underestimated individual harvest site areas considerably before 2012, which was not the case after 2012. This can to some degree explain the smaller efficiency of MA estimators for 2014 at least in Norway because FCL data from 2010-2014 were used in this estimate. Alternatives to FCL may be LandTrendr (Kennedy et al. 2018), which is also based on Landsat timeseries, or Sentinel-2 based algorithms (Mikeladze et al. 2020). The latter may become of interest for GHGIs once a longer time series and large-scale products based on Sentinel-2 become available.

Average annual MA estimators were considerably more efficient than pooled estimators due to three reasons. First, all considered counties except for Denmark started with national ALS campaigns after 2009. Due to our data requirements, ALS data could therefore not be used in the pooled estimates for most countries. Second, because the NFI data used here are based on 5 annual panels measured in 2014-2018 covering a period of 5 years each, the FCL data have to cover a period of 10 years for the pooled estimator. This can result in a temporal mismatch: For example, a clear cut may have been observed by FCL in 2010 for a plot belonging to the 2018 panel. This C-stock change would have been recorded by the NFI in 2013. However, the recorded C-stock loss of 2013 does not belong to the 2014-2018 NFI panels. Instead, in 2018, no further C-stock loss would likely be recorded and hence the FCL would be considered a commission error. Third, trends in C-stock losses, as they are clearly visible e.g. for Norway, are reflected in changes of parameter estimates of annual working models. The residuals of annual working models can therefore be smaller than those of a model for the pooled data, which reduces the variance of the average annual MA estimates compared to the pooled MA estimates.

Climate-change mitigation efforts such as afforestation, fertilization, re-wetting, or reduction of deforestation, are typically planned on the level of counties or municipalities in the Nordic countries. Therefore, information needs with respect to C-stock changes have increased on sub-national scales in the recent years (NEPA 2020). Due to the smaller number of NFI plots available, the use of FCL and ALS data allowed even larger efficiency gains at these scales than at the national level. Similar results have been reported also for other response variables in combination with various remotely sensed data (e.g. Breidenbach and Astrup 2012, Pulkkinen et al. 2018, Haakana et al. 2020).

In GHGIs, C-stock changes need to be reported for six main land-use categories including forest land, grasslands, and wetlands, but also for all change categories among them (IPCC 2006). In addition, these land-use categories must be reported separately for mineral and organic soils. While our proposed concept in general is capable of supporting these estimates, additional information may be required to ensure MA estimators that are more efficient than BE estimators, which merely utilize the NFI field data. For estimates of land-use categories like grasslands or change categories such as afforestation or deforestation, additional maps that accurately describe these categories would be required to enable the use of models that are specific for these categories. While these maps may be available in some countries, they often exhibit low accuracies for changes and they only start being available for larger regions where their accuracy may not yet be sufficient (EEA 2019).

The reason why the efficiency of our estimators was very similar for forest lands as for all land-use categories was that the clear majority of living biomass C-stock losses in Northern Europe comes from harvests in remaining forests. Therefore, the proposed approach could already now be useful in GHGIs.

# 6 Conclusions

The following conclusions can be drawn from this study. i) The use of the global Landsat-based product FCL in combination with NFI data improved estimates of C-stock losses. ii) The additional use of ALS data to describe the C-stock before a canopy loss tended to further improve estimates but, in some cases, also reduced the precision of estimates compared to using FCL only. iii) Efficiency gains, in terms of variance ratios, of more than 2 on national scales were observed when utilizing remotely-sensed data in addition to NFI field plots. Even larger efficiency gains can be obtained on sub-national scales. iv) Due to their better temporal match, average annual C-stock change estimates resulted in more precise estimates than pooled estimates when used in combination with remotely-sensed data.

# 7 Acknowledgements

We acknowledge the provision of field data by the National Forest Inventories of Norway, Sweden, Denmark, and Latvia. We would like to thank Johannes Schumacher (NIBIO), Marius Hauglin (NIBIO), Jörgen Wallerman (SLU), Torben Riis-Nielsen (University of Copenhagen), and Jurgis Jansons (Silava) for supporting the data generation. This study was funded through the FACCE ERA-GAS project INVENT (NO: NRC 276398, SE: FORMAS FR-2017/0006, DK: DIF 7108-00003b, LV: ES RTD/2017/32). The cooperation with Annika Kangas was supported by the SNS-funded CARISMA network.

# 8   Appendix

## 8.1   Synthetic estimates

The total area of a country or stratum is denoted $\lambda$. For the MA estimator using the FCL model [model (11)], the area with forest canopy changes ($FCL = 1$) is $\lambda_{CL}$, whereas the remaining area is $\lambda_N$. It follows that $\lambda = \lambda_{CL} + \lambda_N$. The synthetic estimator in the area with FCL=1 is

$$\hat{t}^S_{CL} = \lambda_{CL}\bar{y}_{CL}. \tag{14}$$

and the synthetic estimator in the remaining area is

$$\hat{t}^S_N = \lambda_N \bar{y}_N. \tag{15}$$

The synthetic C-stock loss estimator that is required in MA estimator (8) is the sum of the two estimators

$$\hat{t}^S = \hat{t}^S_N + \hat{t}^S_{CL}. \tag{16}$$

For the MA estimator that utilizes ALS-FCL model [model (12)], the area with ALS coverage (ALS-year $\leq [t-5]$) and forest canopy changes ($FCL = 1$) is denoted $\lambda_L$, the area without ALS coverage but with forest canopy changes ($FCL = 1$) is $\lambda_{CL}$, whereas the remaining area is $\lambda_N$. It follows that $\lambda = \lambda_{CL} + \lambda_L + \lambda_N$. The population-level ALS mean height above ground within areas with forest canopy changes ($FCL = 1$) is denoted $\bar{X}_L$. That is, $\bar{X}_L$ is the mean height over all ALS first returns covered by an FCL=1 pixel. The synthetic C-stock loss estimator within ALS coverage is therefore

$$\hat{t}^S_L = \lambda_L \times (\hat{\beta}_0 + \hat{\beta}_1 \bar{X}_L + \hat{\beta}_2 \bar{X}_L^2). \tag{17}$$

The synthetic estimators $\hat{t}^S_{CL}$ and $\hat{t}^S_N$ are calculated as in eqs. (14) and (15) but with different total areas. The synthetic C-stock loss estimator that is required in MA estimator (8) is the sum of three estimators

$$\hat{t}^S = \hat{t}^S_L + \hat{t}^S_N + \hat{t}^S_{CL}. \tag{18}$$